\documentclass[12pt]{article}
\usepackage{hhline}
\usepackage{latexsym,graphicx}
\usepackage{subfigure}
\usepackage{amssymb}
\usepackage{amsmath}
\usepackage{amscd}
\usepackage{amsthm}
\usepackage{float}
\usepackage{xcolor}
\usepackage[left=2cm,top=2.5cm,right=2.5cm,bottom=1.5cm]{geometry}   
 
\begin{document}
\begin{center}
\large{\bf{Probing $\kappa(R,T)$ cosmology via empirical approach}} \\
\vspace{10mm}
\normalsize{Nasr Ahmed$^1$,$^2$ and Anirudh Pradhan$^3$}\\
\vspace{5mm}
\small{\footnotesize $^1$ Mathematics Department, Faculty of Science, Taibah University, Saudi Arabia.} \\
\small{\footnotesize $^2$ Astronomy Department, National Research Institute of Astronomy and Geophysics, Helwan, Cairo, Egypt\footnote{nasr.ahmed@nriag.sci.eg}}\\
\small{\footnotesize $^3$ Department of Mathematics, Institute of Applied Sciences \& Humanities,
GLA University, Mathura-281 406, Uttar Pradesh, India \footnote{pradhan.anirudh@gmail.com}}
\end{center}   
\date{}
\begin{abstract}
In this paper, a stable flat cosmological model has been constructed and the evolution of dark energy has been investigated in the framework of the recently suggested $\kappa(R,T)$ gravity. The empirical approach we adopt in the current work reveals some interesting cosmological features consistent with observations and the standard $\Lambda$CDM model. The evolution of cosmic pressure shows a positive-to-negative transition corresponding to the cosmic deceleration-acceleration transition, both the deceleration parameter and the cosmic pressure have the positive-to-negative sign flipping. While this behavior can provide explanation for the deceleration-acceleration transition, the reason behind this positive-negative transition itself is still missing.

\end{abstract}

\section{Introduction}

A very interesting topic in modern cosmology is the accelerating cosmic expansion \cite{11,13,14}. This cosmic acceleration has been a major motivation behind introducing a wide range of modified gravity theories in order to find a satisfactory explanation to this problem. An exotic form of energy with negative pressure, named dark energy, has been assumed where its negative pressure plays the role a repulsive gravity forcing the expansion to accelerate. Several dark energy models have been suggested constructed mainly through two approaches: dynamical scalar fields \cite{quint}-\cite{ark22} and modified gravity theories \cite{noj1}-\cite{nass22}. In the current work, we are interested in a recently suggested Non-Lagrangian modified gravity theory based on a natural generalization of Einstein equations \cite{basic}. While the least action principle represents a major tool to build a physical theory, the gravitational theory introduced in \cite{basic} is not based on this principle. The modified Einstein equations are given as 
\begin{equation}
G_{\mu\nu}-\Lambda g_{\mu\nu}=\kappa(R,T) T_{\mu\nu},
\end{equation}
Where $G_{\mu\nu}=R_{\mu\nu}-\frac{1}{2}g_{\mu\nu}R$ is the Einstein tensor, $\Lambda$ is the cosmological constant, $g_{\mu\nu}$ is the metric tensor, $T_{\mu\nu}$ is the energy-momentum tensor. The function $\kappa(R,T)$ is a generalization of the Einstein gravitational constant which allows exploring the possibility of a running gravitational constant. This approach implies the non-covariant conservation of the energy-momentum tensor which happens in other modified gravity theories as well such as Rastall gravity \cite{rastall} and $f(R,T)$ gravity \cite{frt}. Several models can be explored according to different choices of the function $\kappa(R,T)$. Two particular choices have been studied with their cosmological consequences in \cite{basic}: $\kappa(T)=8 \pi G-\lambda T$ which corresponds to a matter-matter coupling, and $\kappa(R)=8 \pi G+\alpha R$ which corresponds to a matter-curvature coupling. The coupling constants $\lambda$ and $\alpha$ have been assumed to be sufficiently small so it can be consistent with a small violation of the covariant
conservation of the energy-momentum tensor. Original general relativity can be recovered when $ \lambda, \alpha \rightarrow 0$. The modified Friedmann equations in this gravity theory are given as:
\begin{equation}\label{cosm3}
\left(\frac{\dot{a}}{a}\right)^2+\frac{1}{K^2a^2}-\frac{\Lambda}{3}=\frac{\kappa(R,T)}{3}\rho(t),
\end{equation}
\begin{equation} \label{cosm4}
\frac{\ddot{a}}{a}= \frac{\Lambda}{3}-\frac{\kappa(R,T)}{6} \left(3p(t)+\rho(t)\right).
\end{equation}
Where $K^{-1}=0$ for a flat universe.

\section{The cosmological model}
\begin{figure}[H]
  \centering    
	\subfigure[$q$]{\label{F636}\includegraphics[width=0.3\textwidth]{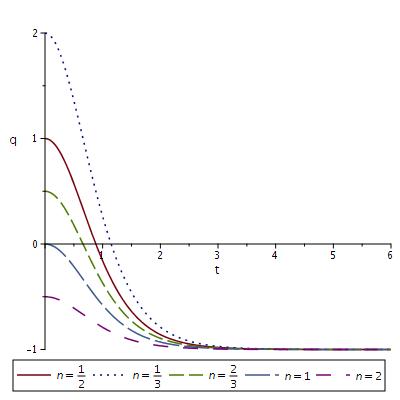}} 
	\subfigure[$j$]{\label{F63655}\includegraphics[width=0.3\textwidth]{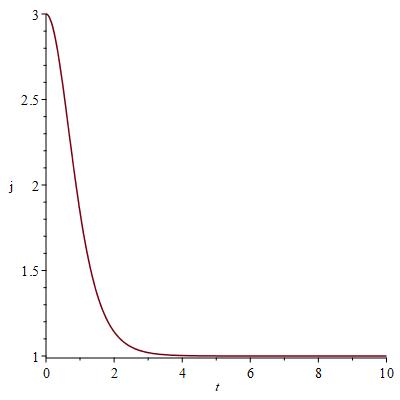}}
		\subfigure[$\rho(t)$]{\label{F6d}\includegraphics[width=0.3\textwidth]{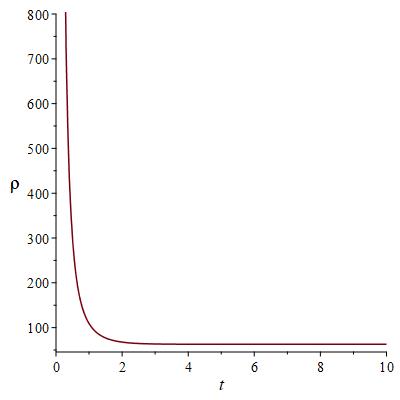}}\\
	\subfigure[$p(t)$]{\label{F6}\includegraphics[width=0.3\textwidth]{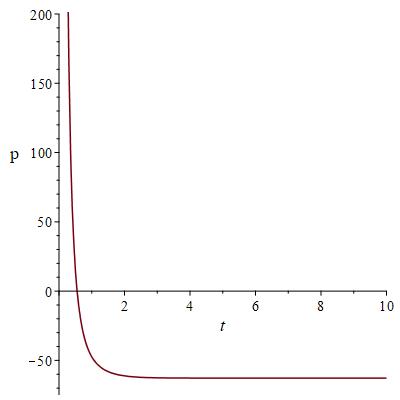}}
	\subfigure[$\omega(t)$]{\label{F6gdd}\includegraphics[width=0.3\textwidth]{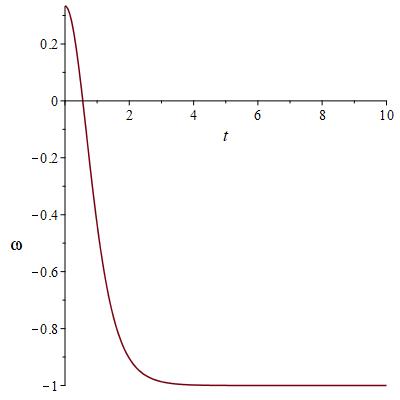}}
	\subfigure[$\omega(z)$]{\label{F6gg}\includegraphics[width=0.3\textwidth]{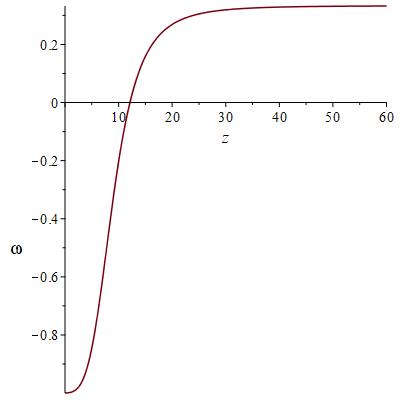}}\\
		\subfigure[Classical EC]{\label{F6rd}\includegraphics[width=0.3\textwidth]{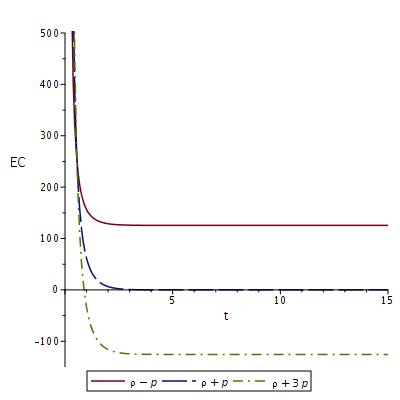}}
			\subfigure[Nonlinear EC]{\label{F6rdg}\includegraphics[width=0.3\textwidth]{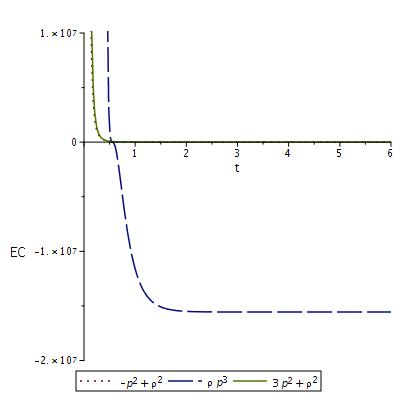}}
				\subfigure[$v_s^2$]{\label{Fdg}\includegraphics[width=0.3\textwidth]{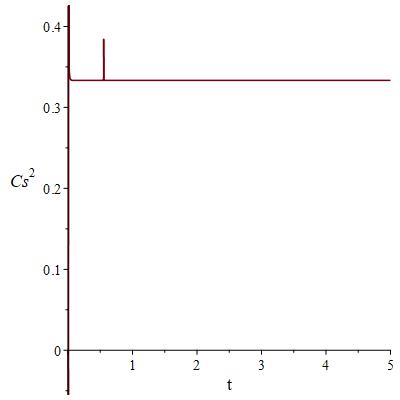}}
  \caption{\ref{F636} The deceleration parameter $q$ flips sign from positive (decelerating phase) to negative (accelerating phase) for $0<n<1$. For $n=\frac{1}{2}$ we get $-1< q < 1$ .\ref{F63655} The jerk parameter has the asymptotic value $j=1$ at late-times where the present model tends to a flat $\Lambda CDM$ model. \ref{F6d} The physical accepted behavior of the energy density. \ref{F6} The cosmic pressure also flips sign from positive in the early time decelerating era to negative in the late time accelerating era. \ref{F6gdd} The evolution of EoS patrameter with cosmic time. \ref{F6gg} The evolution of EoS parameter with the redshift where $\omega \rightarrow -1$ as $z \rightarrow 0 $. The rest of figures show the stability conditions. Here we have chosen $n=\frac{1}{2}$, $\lambda=0.1$, $\Lambda=0.01$, $A=0.1$ and $\xi = 1$.}
  \label{fig:cassimir5cc5}
\end{figure}

Considering an homogeneous and isotropic universe filled by a perfect fluid, the energy-momentum tensor is
\begin{equation}
T_{\mu\nu}=(p+\rho)u_{\mu}u_{\nu}-pg_{\mu\nu},
\end{equation}
Where $p$, $\rho$ and $u_{\mu}$ are the pressure, density and the 4 velocity vector respectively. The FLRW metric given by
\begin{equation}
ds^{2}=-dt^{2}+a^{2}(t)\left[ \frac{dr^{2}}{1-Kr^2}+r^2d\theta^2+r^2\sin^2\theta d\phi^2 \right] \label{RW}
\end{equation} 
where $r$, $\theta$, $\phi$ are comoving spatial coordinates, $a(t)$ is the cosmic scale factor, $t$ is time, $K$ is either $0$, $-1$ or $+1$ for flat, open and closed universe respectively. Considering only the spatially flat case ($K=0$) suggested by observations \cite{flat1, flat2, flat3}, and choosing the function $\kappa(R,T)$ as $\kappa(T)=8\pi G-\lambda T$ the modified Friedmann equations become \cite{basic}:
\begin{equation}\label{cosm1}
\left(\frac{\dot{a}}{a}\right)^2=\frac{8\pi G}{3}\rho+\frac{\Lambda}{3}-\frac{\lambda \rho}{3} (\rho-3p),
\end{equation}
\begin{equation} \label{cosm2}
\frac{\ddot{a}}{a}=-\frac{4\pi G}{3}(3p+\rho)+\frac{\Lambda}{3}+\frac{\lambda}{6}(\rho-3p)(\rho+3p),
\end{equation}
We utilize the following empirical form which produces a deceleration-acceleration cosmic transition for $0 < n < 1$ where the associated deceleration parameter $q(t)$ flips sign from positive to negative ( Figure \ref{fig:cassimir5cc5}(a)), 
\begin{equation} \label{ansatz}
a(t)= A\sinh^n(\xi t)
\end{equation}
We get the deceleration and jerk parameters respectively as
\begin{equation} \label{q1}
q(t)=-\frac{\ddot{a}a}{\dot{a}^2}=\frac{-n\cosh^2(\xi t)+1}{n\cosh^2(\xi t)}
\end{equation}
\begin{equation}
j(t)=\frac{\dddot{a}a^2}{\dot{a}^3}=\frac{-n^2\cosh^2(\xi t)+3n-2}{n^2\cosh^2(\xi t)}
\end{equation}
A possible way to describe models close to $\Lambda$CDM is by using the jerk parameter which has the value $j=1$ for flat $\Lambda$CDM models \cite{jerk1,jerk2,81}. For the current flat model, we can see in Figure \ref{fig:cassimir5cc5}(b) that this parameter has the asymptotic value $j=1$ at late-time. So, in addition to the observationally accepted behavior of $q(t)$, the evolution of $j(t)$ provides another support for the ansatz (\ref{ansatz}). This hyperbolic solution also appears in different cosmological contexts where good agreement with observations has been obtained such as Bianchi cosmological models \cite{pr}, quintessence cosmology \cite{sen}, scalar field cosmology in modified $f(R)$ gravity \cite{senta}, evolution of dark energy in Chern-Simons modified gravity \cite{sent}. It has been shown in \cite{nas2} that a stable flat entropy-corrected cosmology can be obtained through this ansatz. It has been stated in \cite{sen} that the main motivation behind using such hyperbolic ansatz is its consistency with observations. Solving (\ref{cosm1}) and (\ref{cosm2}) using (\ref{ansatz}), we get the expressions for the cosmic pressure and density as

\begin{equation} \label{p}
p(t)= \frac{5\, \left(  \left( {\xi}^{2}-\frac{4}{5}\,\Lambda \right) {{\rm e}^{2\,\xi t}}-\frac{3}{10}
\, \left( {{\rm e}^{4\,\xi t}}+1 \right)  \left( {\xi}^{2}-\frac{4}{3}\,\Lambda
 \right)  \right)  \left( 4\,\pi+\sqrt {\lambda\, \left( -3\,{\xi}^{2}+4
\,\Lambda \right) +16\,{\pi}^{2}} \right) 
}{6\,\lambda\, \left( {\xi}^{2}-\frac{4}{3}\,\Lambda \right)  \left( {{\rm e}^{\xi t}
}+1 \right) ^{2} \left( {{\rm e}^{\xi t}}-1 \right) ^{2}
}
\end{equation}

\begin{equation} \label{rho}
\rho(t)=\frac{(3\,\xi ^2l^2-4\,l^2\Lambda+4\Lambda)(32\,\pi p(t)\,l^2+3\,\xi^2l^2-4\,l^2\Lambda-32\,\pi p(t)-4\,\xi^2+4\Lambda)}{48\,\lambda \,p(t)\,(l-1)^2\,(l+1)^2\,(\xi^2-\frac{4}{3}\Lambda)}
\end{equation}
Where $l=\cosh(\xi t)$. The expression for the equation of state parameter is obtained directly as $\omega=\frac{p(t)}{\rho(t)}$. Figure \ref{fig:cassimir5cc5}(c) shows the physically accepted behavior of the energy density where $\rho(t) \rightarrow \infty$ as $t \rightarrow 0$. The evolution of cosmic pressure (Figure \ref{fig:cassimir5cc5}(d)) shows a sign flipping from positive in the early-time decelerating epoch to negative in the late-time accelerating epoch without "`pressure singularity"'. Such behavior of cosmic pressure in the current $\kappa(R,T)$ gravity cosmological model is consistent with the "`dark energy"' DE assumption where the negative pressure acts as a repulsive gravity forces the expansion to accelerate. It is also believed that the early universe was dominated by positive pressure where the cosmic expansion was decelerating. So, both the deceleration parameter and the cosmic pressure have the positive-to-negative sign flipping. Figure \ref{fig:cassimir5cc5}(e) shows that the evolution of the equation of state parameter EoS with cosmic time is a decreasing fuction restricted to the range $-1< q < \frac{1}{3}$. It starts from $\simeq \frac{1}{3}$ (radiation-like epoch) and then crosses the zero (dust epoch $\omega = 0$) to the negative domain. It finally reaches the current dark energy-dominated epoch at $-1$ without crossing this cosmological constant boundary (phantom divide line) at $\omega=-1$ to the phantom era ($\omega<-1$), i.e. no Quintom behavior. Quintom is a dynamical model of dark energy where $\omega(t)$ can smoothly cross over $\omega=-1$ which is favored by observations \cite{quintom}. Since observations show that $\omega$ is roughly equal to $-1$, the current accelerated universe could be in the cosmological constant era ($\omega=-1$), quintessence era ($-1<\omega<-1/3$) or phantom era ($\omega<-1$). Many authors investigated whether DE can evolve to the phantom era or not. It has been argued in \cite{vikman} that if the DE transition from $\omega \geq -1$ in the close past to $\omega <-1$ at the current era has been confirmed by observations, then such transition can not be explained by the classical dynamics described by an effective scalar field Lagrangian. This transition can be allowed when more complicated physics is considered \cite{d1,d2,d3}. The EoS parameter $\omega$ is equal to $-1$ for the current dark energy-dominated epoch with redshift $z\simeq 0$. Using $a=\frac{1}{1+z}$, we obtain the following expression for $\omega(z)$
\begin{equation}
\omega(z)=-\frac{128}{9}\frac{ f(z)}{g(z)},~~~~~~~~~~~~~~~~~~~~~~~~~~~~~~~~~~~~~~~~~~~~~~~~~~~
\end{equation}
Where 
\begin{eqnarray*}
f(z)&=&(1+\frac{1}{2}r^2+r~l)~r^4(l-1)^2(\xi^2r^2-3\xi^2+4\Lambda)(l+1)^2 \times \\
&&(\pi+\frac{1}{4}\sqrt{\lambda(-3\xi^2+4\Lambda)+16\pi^2})^2,\\ 
g(z)&=& (\xi^2r^2+\xi^2-\frac{4}{2}\Lambda)(1-r(l-1))^2(1+r(l+1))^2 \times \\  
&&(\xi^2-\frac{32}{3}\pi^2-\frac{8}{3}\pi\sqrt{\lambda(-3\xi^2+4\Lambda)+16\pi^2}-\frac{4}{3}\Lambda \lambda),
\end{eqnarray*}
Where $r=A^2(1+z)^2$ and $l=\sqrt{\left(\frac{1+r^2}{r^2}\right)}$. Figure \ref{fig:cassimir5cc5}(e) shows that for the current model $\omega(z \simeq 0 ) \simeq -1$ for $\xi=1$, $A=0.1$, $\lambda=0.1$ and $\Lambda=0.01$. So, we have used these values in all plots. We can also see that for the early Universe ($z \rightarrow \infty$) we get $\omega(z \rightarrow \infty) \simeq \frac{1}{3}$ i.e. a radiation dominated universe which is the same analysis we get from the evolution of $\omega(t)$. 
		
\section{Stability of the model}

In this section we test the stability of the current through the classical linear energy conditions \cite{ec11,ec12} and the sound speed. In addition, and because of the existence of quadratic energy density terms in the cosmological equations, we also test the validity of the ew nonlinear energy conditions (ECs) \cite{ec,FEC1, FEC2, detec}. The classical linear ECs (namely, the null $\rho + p\geq 0$; weak $\rho \geq 0$, $\rho + p\geq 0$; strong $\rho + 3p\geq 0$ and dominant $\rho \geq \left|p\right|$ energy conditions) should be replaced by other nonlinear ECs in the presence of semi-classical quantum effects \cite{ec, detec}. Moreover, in completely general situations the ECs can not be still valid and then such conditions can't be considered as fundamental physics \cite{ec2,parc}. The nonlinear ECs we consider here are (i) The flux energy condition (FEC): $\rho^2 \geq p_i^2$ \cite{FEC1, FEC2}. (ii) The determinant energy condition (DETEC): $ \rho . \Pi p_i \geq 0$ \cite{detec}. (iii) The trace-of-square energy condition (TOSEC): $\rho^2 + \sum p_i^2 \geq 0$ \cite{detec}. \par

The strong EC ($\rho + 3p\geq 0$) represents a strong restriction on gravity where 'it should always be attractive'. However, it has been indicated that such strong EC fails even in the classical regime when describing the current cosmic accelerating era (assumed to be dominated by negative pressure which represents a repulsive gravity) and during inflation \cite{ec3,ec4,ec5}. Since negative pressure dominates the late-time epoch in the present model, we don't expect the strong EC to be satisfied in the late-time era. This can be seen in Figure 1(g) where the strong EC is invalid during the late-time DE dominated era. The dominant EC represents the condition that energy density should be non-negative, Figure 1(g) shows that this condition is always satisfied. The same figure also shows that the weak EC is also valid all the time. The nonlinear ECs have been plotted in Figure 1(h) which shows that the flux EC and the trace-of-square EC are always valid all the time while the determinant EC is violated in late-time. Since causality requires that sound speed $v$ must be less than the speed of light $c$, then in the relativistic units ($c=G=1$) the condition $0 \leq \frac{dp}{d\rho} \leq 1$ should be always satisfied where $\frac{dp}{d\rho}=v_s^2$ is the adiabatic square sound speed. For the current model we have
\begin{equation}
v_s^2=-\frac{f_1(t)}{13 (e^{\xi t}-1)^3(e^{\xi t}+1)^3g_1(t)}.
\end{equation}
Where 
\begin{eqnarray}
f_1(t)&=&\left( (-\frac{10}{3}\xi^2+\frac{8}{3}\Lambda) e^{2\xi t}+(\xi^2-\frac{4}{3}\Lambda)(e^{4\xi t}+1) \right)^2(\cosh(\xi t)+1)^3(\cosh(\xi t)-1)^3e^{2\xi t} \\ \nonumber
&\times& (4\pi+\sqrt{\lambda(-3\xi ^2+4\Lambda)+16\pi^2})^2(e^{2\xi t}+1),
\end{eqnarray}
\begin{eqnarray*}
g_1(t)=-\frac{236}{39}\left(-\frac{30}{59}(\xi^2-\frac{4}{5}\Lambda)(\xi^2-\frac{4}{3}\Lambda)(e^{2\xi t}-e^{6\xi t})+(b^4-\frac{104}{59}\Lambda \xi^2+\frac{48}{59}\Lambda^2)e^{4\xi t}+\frac{9}{118}(\xi^2-\frac{4}{3}\Lambda)^2(e^{8\xi t+1})\right)\\   \nonumber
\times \pi\, \left( l +1 \right) l  \left( l -1 \right) s \sqrt {\lambda\, \left( -3\,{\xi}^{2}+4\,\Lambda \right) +16\,{
\pi}^{2}} -\frac{3}{13} (\xi^2-\frac{4}{3}\Lambda)e^{2\xi t} \times~~~~~~~~~~~~~~~~~~~~~~~~~~~~~~~~~~~~~~~\\   \nonumber
 \left( \lambda\, ( {\xi}^{2}-\frac{4}{3}\,\Lambda ) ^{2} l^{6}-\frac{7}{3}\,\lambda\, ( {\xi}^{2}-\frac{4}{3}\,
\Lambda )  ( {\xi}^{2}-{\frac {12\,\Lambda}{7}} ) 
  l^4 + \frac{8}{3}\, ( \frac{4}{3}\,{\Lambda}^{2}\lambda+ ( -\frac{7}{3}\,{\xi}^{2}\lambda+16
\,{\pi}^{2} ) \Lambda+{\xi}^{4}\lambda-20\,{\pi}^{2}{\xi}^{2}
 ) s l^3  \right.  \\    \nonumber
\left.  + \frac{4}{3}\,\lambda\, ( {\xi}^{4}-\frac{14}{3}\,\Lambda\,{\xi}^{2}+4\,{\Lambda}^{2}
 )  l^2 -  \frac{16}{3}\, ( \frac{2}{3}\,{\Lambda}^{2}\lambda+ ( -\frac{5}{3}\,{\xi}^{2}\lambda+8
\,{\pi}^{2} ) \Lambda+{\xi}^{4}\lambda-10\,{\pi}^{2}{\xi}^{2}
 ) s l +{\frac {16\,
\Lambda\,\lambda\, ( {\xi}^{2}-\Lambda ) }{9}} \right) \\    \nonumber
+\frac{3}{13} \left( \lambda\, ( {\xi}^{2}-\frac{4}{3}\,\Lambda ) ^{2} l^{6}-\frac{7}{3}\,\lambda\, ( {\xi}^{2}-\frac{4}{3}\,
\Lambda )  ( {\xi}^{2}-{\frac {12\,\Lambda}{7}} ) l^{4}-\frac{8}{3}\, ( \frac{4}{3}\,{\Lambda}^{2}\lambda+ ( -\frac{7}{3}\,{\xi}^{2}\lambda+ 16\,{\pi}^{2} ) \Lambda+{\xi}^{4}\lambda-20\,{\pi}^{2}{\xi}^{2}
 ) s  l^{3}\right.  \\    \nonumber
\left. \frac{4}{3}\,\lambda\, ( {\xi}^{4}-\frac{14}{3}\,\Lambda\,{\xi}^{2}+4\,{\Lambda}^{2}
 )  l^{2}+\frac{16}{3}\, ( 2
/3\,{\Lambda}^{2}\lambda+ ( -\frac{5}{3}\,{\xi}^{2}\lambda+8\,{\pi}^{2}
 ) \Lambda+{\xi}^{4}\lambda-10\,{\pi}^{2}{\xi}^{2} ) s l +{\frac {16\,\Lambda\,
\lambda\, ( {\xi}^{2}-\Lambda ) }{9}}
\right)\\ \nonumber
\times  \left( {\xi}^{2}-\frac{4}{3}\,\Lambda \right) {{\rm e}^{6\,\xi t}}+ sl\left(  e^{\xi t} ( ( -{\frac {64\,{\Lambda}^{3}\lambda}{39}}+ ( {\frac {496\,{
\xi}^{2}\lambda}{117}}-{\frac {256\,{\pi}^{2}}{13}} ) {\Lambda}^{2
}+ ( -{\frac {140\,{\xi}^{4}\lambda}{39}}+{\frac {128\,{\pi}^{2}{\xi}
^{2}}{3}} ) \Lambda+~~~~~~\right.  \\    \nonumber
{\xi}^{6}\lambda-{\frac {944\,{\xi}^{4}{\pi}^{2}
}{39}} )  l^{2} 
\left. +{\frac { 64\,{\Lambda}^{3}\lambda}{39}}+ ( {\frac {256\,{\pi}^{2}}{13}}-{
\frac {640\,{\xi}^{2}\lambda}{117}} ) {\Lambda}^{2}+ ( {
\frac {76\,{\xi}^{4}\lambda}{13}}-{\frac {128\,{\pi}^{2}{\xi}^{2}}{3}}
 ) \Lambda-2\,{\xi}^{6}\lambda+{\frac {944\,{\xi}^{4}{\pi}^{2}}{39}}) \right.  \\    \nonumber
\left. + \frac{1}{26} { ( 3\, ( {\xi}^{2}\lambda-16\,{\pi}^{2}-\frac{4}{3}\,\Lambda\,
\lambda )  l^{2}-6\,{\xi}^{2}\lambda+48\,{\pi}^{2}+4\,\Lambda\,\lambda )  ( {\xi}^{2}-\frac{4}{3}\,\Lambda ) ^{2} ( {{\rm e}^{8\,\xi t}}+1 ) } \right)~~~~~~~~~~~~~~~~~~~~~~~~~~~~~~
\end{eqnarray*}
Where $l=\cosh ( \xi t )$ and $s=\sinh ( \xi t )$. Figure 1(i) shows that this stability condition is satisfied in early and late time.

\section{Conclusion}
$\kappa(R,T)$ is a recently suggested Non-Lagrangian modified theory of gravity based on a natural generalization of Einstein equations. In this paper, we have adopted an empirical approach and constructed a stable flat cosmological model in agood agreement with observations. A basic interesting feature in the current model is the sign flipping in the behavior of cosmic pressure from positive in the early decelerating epoch to negative in the late accelerating epoch which agrees with the dark energy assumption. The evolution of dark energy with cosmic time and with redshift has been studied where there is no crossing to the cosmological constant boundary. The stability of the model has been tested through classical energy conditions, non-linear energy conditions, and the sound speed.


\begin{thebibliography}{paper}
\bibitem{11} S. Perlmutter et al., Measurements of Omega and Lambda from 42 High-Redshift Supernovae
,Astrophys. J. 517, 565 (1999).
\bibitem{13} W. J. Percival et al., The 2dF Galaxy Redshift Survey: The power spectrum and the matter content of the universe
, Mon. Not. Roy. Astron. Soc. 327, 1297 (2001).
\bibitem{14} D. Stern et al., straining the Equation of State of Dark Energy. $I: H(z)$ Measurements, J. Cosm. Astrop. Phys. 008, (2010).
\bibitem{quint} S. Tsujikawa, Quintessence: A Review,
Class. Quant. Grav. 30, 214003 (2013).
\bibitem{chap} A. Y. Kamenshchik, U. Moschella \& V. Pasquier, An alternative to quintessence, An alternative to quintessence,
 Phys. Lett. B511, 265 (2001).
\bibitem{phant} R. R. Caldwell, A Phantom Menace? Cosmological consequences of a dark energy component with super-negative equation of state
, Phys. Lett. B 545, 23 (2002).
\bibitem{ess} T. Chiba, T. Okabe \& M. Yamaguchi, Kinetically driven quintessence
, Phys. Rev. D62, 023511 (2000).
\bibitem{tak} A. Sen, Tachyon Matter, JHEP 0207, 065 (2002). 
\bibitem{ark} N. Arkani-Hamed, H. Cheng, M. A. Luty \& S. Mukohyama, Ghost Condensation and a Consistent Infrared Modification of Gravity
, JHEP 0405, 074 (2004).
\bibitem{ark22}Nasr Ahmed \& Sultan Z. Alamri, A cyclic universe with varying cosmological constant in $f(R, T)$ gravity
Can. J. Phys. (2019)  https://doi.org/10.1139/cjp-2018-0635; Nasr Ahmed, Ricci-Gauss-Bonnet holographic dark energy in Chern-Simons modified gravity. 2019, in press, Modern Physics Letters A. arXiv:1901.04849 [gr-qc].
\bibitem{noj1} S. Nojiri \& s. D. Odintsov, Modified $f(R, T)$ gravity consistent with realistic cosmology: From a matter dominated epoch to a dark energy universe, Phys. Rev. D74, 086005 (2006).
\bibitem{noj8} S. Nojiri, S. D. Odintsov \& P. V. Tretyakov, From Inflation to Dark Energy in the Non-Minimal Modified
Gravity, Prog. Theor. Phys. Suppl. 172, 81 (2008).
\bibitem{torsion} R. Ferraro \& F. Fiorini, Modified teleparallel gravity: Inflation without an inflaton
, Phys. Rev. D 75, 084031 (2007).
\bibitem{beng1} G. R. Bengochea \& R. Ferraro, Dark torsion as the cosmic speed-up, Phys. Rev. D 79, 124019 (2009).
\bibitem{de1} A. De Felice \& S. Tsujikawa, $f(R)$ theories, Living Rev. Rel. 13, 3 (2010).
\bibitem{alt1} M. E. S. Alves, O. D. Miranda \& J. C. N. de Araujo, Can massive gravitons be an alternative to dark energy?, Physics Letters B 700 (5), (2011).
\bibitem{alt2} A. Maeder, Dynamical Effects of the Scale Invariance of the Empty Space: The Fall of Dark Matter?, The Astro. J. Vol 849, 2 (2017).
\bibitem{alt3} J. Gagnon and J. Lesgourgues, Dark goo: bulk viscosity as an alternative to dark energy
, J. Cosmol. Astropart. Phys. 09 (026), 2011 [1107.1503]
\bibitem{nass} Nasr Ahmed \& I. G. Moss, Gaugino condensation in an improved heterotic M-theory, JHEP 12, 108 (2008).
\bibitem{nass22} Nasr Ahmed \& I. G. Moss, Balancing the vacuum energy in heterotic M-theory, Nucl. Phys. B 833,1-2 (2010).
\bibitem{basic} Teruel, $\kappa(R,T)$ gravity, G.R.P. Eur. Phys. J. C (2018) 78: 660.
\bibitem{rastall} Rastall,P. "Generalization of the Einstein theory", Phys. Rev. D6 (1972) 3357-3359.
\bibitem{frt} T. Harko et. al., $f(R,T)$ gravity, Phys.Rev.D84:024020 (2011).
\bibitem{flat1} C. L. Bennett et al., First Year Wilkinson Microwave Anisotropy Probe (WMAP) Observations: Preliminary Maps and Basic Results ,Astrophys. J. Suppl. 148, 1 (2003).
\bibitem{flat2} D. N. Spergel et al., First Year Wilkinson Microwave Anisotropy Probe (WMAP) Observations: Determination of Cosmological Parameters, Astrophys. J. Suppl. 148, 175 (2003a).
\bibitem{flat3} Nasr Ahmed \& Sultan Z. Alamri, A stable flat universe with variable Cosmological constant in $f(R,T)$ Gravity
, Res. Astron. Astrophys., Vol 18, No. 10 (2018).
\bibitem{80} Y. Caia, T. Qiua, Y. Piaob et. al., Bouncing Universe with Quintom Matter
, JHEP 0710: 071 (2007).
\bibitem{jerk1} T. Chiba and T. Nakamura, The Luminosity Distance, the Equation of State, and the Geometry of the Universe
, Prog. Theor. Phys. 100, 1077 (1998); V. Sahni, astroph/0211084 (2004); R. D. Blandford, M. Amin, E. A. Baltz, K. Mandel, and P. J.
Marshall, Cosmokinetics, astro-ph/0408279 (2004).
\bibitem{jerk2} M. Visser, Jerk, snap, and the cosmological equation of state
,Class. Quantum Grav. 21, 2603 (2004); M. Visser, Cosmography: Cosmology without the Einstein equations, Gen. Relativ. Gravit. 37, 1541 (2005).
\bibitem{81} D. Rapetti, S. W. Allen, M. A. Amin, and R. D. Blandford, A kinematical approach to dark energy studies
, Mon.Not.Roy.Astron.Soc.375:1510-1520 (2007).
\bibitem{quintom} G.-B. Zhao, J.-Q. Xia, H. Li, C. Tao et. al., Probing for dynamics of dark energy and curvature of universe
with latest cosmological observations, Phys. Lett. B 648: 8-13 (2007).
\bibitem{vikman} A. Vikman, Can dark energy evolve to the Phantom?, Phys.Rev. D71, 023515 (2005).
\bibitem{d1} C. Armendáriz-Pión, Could dark energy be vector-like?, JCAP 0407: 007 (2004).
\bibitem{d2} Bo Feng, Xiulian Wang, Xinmin Zhang, Dark Energy Constraints from the Cosmic Age and Supernova, Phys.Lett.B607:35-41 (2005).
\bibitem{d3} Varun Sahni, Yuri Shtanov, Braneworld models of dark energy, JCAP 0311, 014 (2003).
\bibitem{pr} Nasr Ahmed \& A. Pradhan, Bianchi Type-V Cosmology in $f (R, T)$ Gravity with $\Lambda(T)$, Int. J. Theor. Phys. 53:289–306 (2014).
\bibitem{sen} A. A. Sen \& S. Sethi, Quintessence Model With Double Exponential Potential, Phys. Lett. B532, 159 (2002).
\bibitem{senta} S. D. Maharaj, R. Goswami, S. V. Chervon \& A. V. Nikolaev, Exact solutions for scalar field cosmology in $f(R)$ gravity, Mod. Phys. Lett. A, 32 (30) 1750164 (2017).
\bibitem{nas2} Nasr Ahmed \& Sultan Z. Alamri,  A stable flat entropy-corrected FRW universe. Int. J. Geom. Methods Mod. Phys.  2018. arXiv:1811.08864 [gr-qc].
\bibitem{sent} J. G. Silva \& A. F. Santos, Ricci dark energy in Chern-Simons modified gravity, Eur Phys J C 73 (2500) 2013.
\bibitem{ec11} S. W. Hawking \& G. F. R. Ellis, The large scale structure of spacetime (Cambridge University Press, England 1973).
\bibitem{ec12} R. M. Wald, General Relativity (University of Chicago Press, Chicago 1984).
\bibitem{ec} P. Martın–Moruno and M. Visser, Semiclassical energy conditions for quantum vacuum states, JHEP 1309, 050 (2013) .
\bibitem{FEC1} G. Abreu, C. Barcel´o \& M. Visser, Entropy bounds in terms of the w parameter, J.High Energy Physics JHEP12, 092 (2011).
\bibitem{FEC2} P. Mart´ın-Moruno \& M. Visser, Classical and quantum flux energy conditions for
quantum vacuum states, Phys. Rev. D 88 (6) 061701 (2013).
\bibitem{detec} P. MartnMoruno and M. Visser, Semiclassical energy conditions for quantum vacuum states, JHEP 1309, 050 (2013).
\bibitem{ec3} M. Visser, Energy conditions in the epoch of galaxy formation
, Science 276, 88 (1997).
\bibitem{ec4} M. Visser, General Relativistic Energy Conditions: The Hubble expansion in the epoch of galaxy formation
, Phys. Rev. D 56, 7578 (1997) .
\bibitem{ec5} M. Visser, Energy conditions and galaxy formation, gr-qc/9710010
\bibitem{ec2} C. Barcel´o and M. Visser, Twilight for the energy conditions?
, Int. J. Mod. Phys. D
\bibitem{parc} C. Barcelo and M. Visser, Twilight for energy conditions, Int. J. Mod. Phys. D 11, 1553
(2002).

\end{thebibliography}
\end{document}